# Fiber-based mid-infrared frequency-swept laser at 50 MScans/s via frequency down-conversion of time-stretched pulses

Makoto Shoshin[1], Takahiro Kageyama[1], Takuma Nakamura[2], Kazuki Hashimoto[2], and Takuro Ideguchi[1,2,*]

[1]Department of Physics, The University of Tokyo, Tokyo 113-0033, Japan

[2]Institute for Photon Science and Technology, The University of Tokyo, Tokyo 113-0033, Japan

*ideguchi@ipst.s.u-tokyo.ac.jp

**Abstract**

Increasing the sweep rate of mid-infrared (MIR) frequency-swept sources offers significant potential for various high-speed spectroscopy-based applications. While continuous-wave frequency-swept lasers have achieved sweep rates up to 1 MHz, a recently demonstrated time-stretched ultrashort pulsed laser has reached a significantly higher sweep rate, up to tens of MHz. However, the previous system relied on a bulky femtosecond optical parametric oscillator and produced only ~30 discrete spectral elements due to the use of a free-space time stretcher. In this work, we present a fiber-based frequency-swept MIR source that utilizes the frequency down-conversion of time-stretched near-infrared pulses, employing a compact mode-locked fiber laser and telecommunication fiber. As a proof-of-concept demonstration, we performed MIR spectroscopy of methane gas around 3.4 μm at a rate of 50 MSpectra/s, capturing 220 spectral elements over a range of 19.0 cm$^{-1}$. This compact and robust high-speed MIR frequency-swept laser system holds the potential for deployment in field applications.

**Introduction**

Frequency-swept mid-infrared (MIR) lasers enable highly sensitive broadband spectroscopy measurements, including molecular absorption spectroscopy and depth-resolved imaging of highly scattering media. Various types of frequency-swept MIR lasers have been developed, such as temperature- or current-controlled distributed feedback (DFB) quantum cascade lasers (QCLs) [1], external cavity QCLs (EC-QCLs) [2], and MIR lasers utilizing acousto-optic tunable filters (AOTF) [3–6]. Increasing the sweep rate is crucial for applications like combustion diagnosis [7], snapshot stand-off sensing [8], and high-speed MIR-OCT [9]. Active scanners have been primarily employed for high-speed sweeping, including EC-QCLs or difference-frequency-generation (DFG) lasers equipped with rapid angle-scanning mirrors/gratings [9–12] or acousto-optic modulators (AOM) [7,8,13]. However, the sweep rates of these lasers are limited by the operational frequency of the active scanners, up to hundreds of kHz. Another approach involves intra-pulse chirping of DFB-QCLs [14], which has demonstrated a sweep range of 1 cm$^{-1}$ at a rate of 1 MHz, the highest rate reported to date to the best of our knowledge [15].

The frequency chirping of ultrashort pulses from a mode-locked laser provides a passive method for sweeping laser frequency, overcoming the speed limitations of conventional techniques. However, pulses stretched with traditional methods, such as grating pairs, typically reach durations of only hundreds of picoseconds, which are too short for use as frequency-swept sources because high-speed electronic detection systems cannot resolve their spectra. This limitation can be mitigated by significantly stretching the pulses to ~10 ns, enabling the resolution of ~100 spectral elements using a detection electronic bandwidth of ~10 GHz. Recently, we demonstrated the stretching of ultrashort MIR pulses to durations exceeding 10 ns using a free-space angular-chirp-enhanced delay (FACED) method and measured the pulses with a 5-GHz detection bandwidth [17]. By employing a synchronously pumped femtosecond optical parametric oscillator (fs-OPO) driven by a Ti:Sapphire mode-locked laser, we achieved an unprecedented sweep rate of 80 MHz. However, this system generated only ~30 discrete spectral elements due to limitations in the FACED time-stretching mechanism. Additionally, the setup is bulky, requiring a fs-OPO pumped by a Ti:Sapphire laser and a free-space time stretcher with custom-made long mirror pairs, making it unsuitable for practical, field-based applications.

In this work, we present a passively scanned, fiber-based MIR frequency-swept source operating at a scan rate of 50 MScans/s. This is achieved through the frequency down-conversion of time-stretched 1.5-μm femtosecond pulses generated by an erbium-doped mode-locked fiber laser. The frequency down-conversion enables time-stretching using telecommunication fiber, providing a low-loss and continuous frequency sweep. Using the developed laser source in conjugation with a quantum cascade detector (QCD) [18], we demonstrated MIR spectroscopy of methane gas over a -20-dB bandwidth of 19.0 cm$^{-1}$ with 220 spectral elements at a rate of 50 MSpectra/s.

**Results**

Figure 1 illustrates a schematic diagram of the developed MIR frequency-swept source. In the initial stage of the system, near-infrared (NIR) ultrashort pulses are temporally stretched using an optical fiber. These pulses, centered at 1.546 μm with a -20-dB bandwidth of 25 cm$^{-1}$ (6.0 nm), are generated by a 50-MHz erbium-doped mode-locked fiber laser (Femtolite, IMRA America Inc.) and spectrally filtered with optical bandpass filters. The NIR pulses are stretched to 20 ns—matching the interval between adjacent pulses—using a dispersion compensating fiber (DCF) (AD-SM-C-120-FC/APC-3C/3B-10, YOFC). The DCF has a length of 20 km and a group velocity dispersion (GVD) of 255.5 ps$^2$ km$^{-1}$. The average input power of the pulses into the DCF was 0.98 mW. To compensate for propagation losses in the long fiber, Raman amplification was applied by bidirectional pumping with a 1.455-μm laser diode. After propagation through the DCF, the stretched pulses, with an average power of 0.13 mW, were amplified to 760 mW using two stages of erbium-doped fiber amplifiers (EDFA100P, Thorlabs, and a homemade device).

The amplified 1.5-μm stretched pulses are subsequently down-converted to 3.4 μm via difference frequency generation (DFG) using a 20-mm-long periodically poled lithium niobate (PPLN) waveguide (WD-3418-000-A-C-C-TEC, NTT Electronics) with a poling period of 28.5 μm. A fiber-coupled 1.064-μm continuous-wave (CW) distributed Bragg reflector laser (PH1064DBR200BF, Photodigm) serves as the counterpart for the DFG process.

The single longitudinal mode CW laser with a 10 MHz bandwidth ensures a precise one-to-one spectral transfer of the frequency-chirped NIR pulses to the MIR pulses. The CW laser is amplified to 980 mW using a homemade ytterbium-doped fiber amplifier and combined with the stretched 1.5-µm pulses through a wavelength division multiplexing coupler. The combined beams are focused onto the PPLN waveguide, generating MIR pulses with an average power of 9.7 mW at a conversion efficiency of 1.3%/W, including the coupling loss. The inset of Fig. 1 displays the MIR spectrum, centered at 2927 cm$^{-1}$ (3.42 µm), measured using a Fourier-transform infrared spectrometer with a resolution of 0.2 cm$^{-1}$. The MIR spectrum bandwidth is 19.0 cm$^{-1}$ at the -20 dB intensity level, constrained by the phase-matching conditions of the PPLN waveguide. Tuning the MIR spectrum wavelength is achieved by temperature control or using PPLN waveguides with different poling periods. Notably, this turnkey system produces an ultrarapid frequency-swept MIR laser source at a rate of 50 MHz, leveraging off-the-shelf, fiber-based devices in the NIR region.

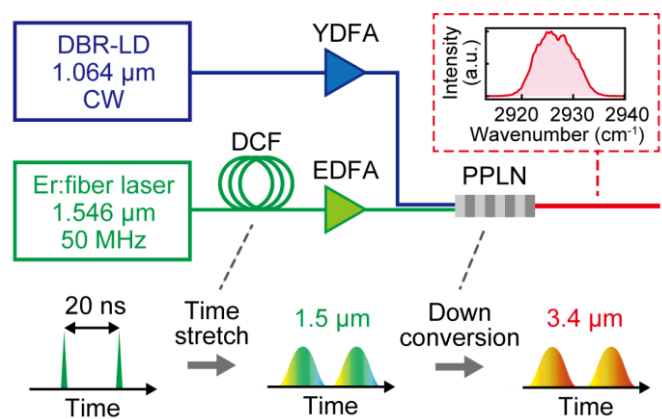

Fig. 1. Schematic of the MIR frequency-swept source via frequency down-conversion of time-stretched NIR pulses. The inset displays the down-converted MIR spectrum on a linear scale, measured using a Fourier-transform infrared spectrometer. DCF, dispersion compensating fiber; EDFA, erbium-doped fiber amplifier; DBR-LD, distributed Bragg reflector laser diode; YDFA, ytterbium-doped fiber amplifier; CW, continuous wave; PPLN, periodically poled lithium niobate.

The high-speed frequency-swept spectra were measured using a high-bandwidth detection system. We employed a QCD with a 10 GHz bandwidth (Hamamatsu Photonics) and an oscilloscope with a 16 GHz bandwidth (WaveMaster 816Zi-B, Teledyne LeCroy). An aspheric lens with a 4.0 mm focal length focused the MIR light onto the QCD. Due to the polarization dependence of the QCD, the polarization of the MIR light was adjusted using a half-wave plate. To address the QCD's low sensitivity (a few mA/W), two-stage microwave amplifiers were utilized. For general measurements, we employed AC-coupled amplifiers with a total gain of 39 dB and a bandwidth of 0.01–10 GHz (EV1HMC8410LP2F, Analog Devices). For the linear interference measurement shown in Fig. 2 (b), DC-coupled amplifiers (117810-HMC460LC5, Analog Devices) with a total gain of 28 dB and a bandwidth of 0–20 GHz were employed.

We first evaluated the instantaneous wavenumber of the stretched MIR pulses. A straightforward method for this evaluation is to measure the beat frequency from heterodyne detection with a CW laser. However, the broad spectral range of our laser (19.0 cm$^{-1}$, corresponding to 570 GHz) necessitates an extremely high-speed detection system for heterodyne measurement. Instead, we used a Michelson interferometer to measure the interference of identical time-stretched MIR pulses with a short delay, providing the time derivative of the instantaneous wavenumber, referred to as the chirp rate. The schematic of this measurement setup is shown in Fig. 2(a). Figure 2(b) presents the interference spectrum averaged over 10,000 measurements. The chirp rate was extracted using the Hilbert transform of this spectrum, as shown in Fig. 2(c). To enhance the signal-to-noise ratio (SNR), a numerical non-causal (zero-phase) bandpass filter with a passband of 2.2-3.6 GHz, which includes the frequency difference between the interfering stretched pulses, was applied before performing the Hilbert transform. The instantaneous wavenumber was then calculated by integrating the curve in Fig. 2(c), as shown in Fig. 2(d). The time difference between the interfering pulses (93.5 ps) was determined by measuring the spectral fringes of the 1.5 μm pulses (Fig. 2(e)), which propagate coaxially with the MIR pulses through the Michelson interferometer. This measurement was performed using an optical spectrum analyzer with a resolution of 0.04 cm$^{-1}$. The absolute wavenumber was calibrated by measuring an absorption line of methane gas, as detailed later. The measured instantaneous wavenumber exhibits a nearly linear sweep, primarily attributed to the group delay dispersion (GDD) of the fiber. A slight nonlinearity, caused by third-order dispersion (TOD), results in an effect within ±0.07 cm$^{-1}$ over a 20-ns time window—smaller than the instrumental spectral resolution (0.086 cm$^{-1}$) of our detection system.

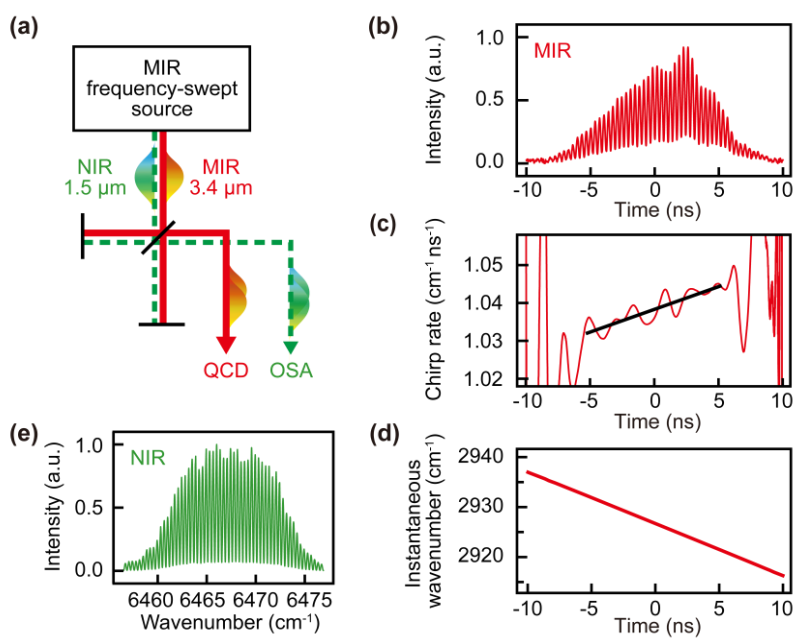

Fig. 2. Evaluation of the instantaneous wavenumber of the stretched MIR pulses. (a) Schematic of the measurement setup using a Michelson interferometer. The MIR and NIR pulses propagate coaxially through the interferometer and are detected separately by a quantum cascade detector (QCD) and an optical spectrum analyzer (OSA). (b) MIR interference spectrum, averaged over 10,000 measurements, captured with the QCD. (c) Chirp rate (time derivative of the instantaneous

wavenumber) extracted from the interference spectrum using the Hilbert transform. The black line represents the fitted curve of the data. (d) Instantaneous wavenumber obtained by integrating the chirp rate. (e) NIR interference spectrum captured with the OSA.

To confirm that the measured instantaneous wavenumber aligns with theoretical expectations, we evaluated the dispersion from the measured data. The instantaneous wavenumber of a stretched pulse within a fiber can be expressed as:

$$\nu(t) = \nu_0 - \frac{1}{2\pi c}\left(\frac{t}{\phi_2} - \frac{\phi_3 t^2}{2\phi_2^3}\right), \tag{1}$$

where $t$, $\nu_0$, $\phi_2$, $\phi_3$ and $c$ represent time, the central wavenumber of the pulse, GDD, TOD, and the speed of light, respectively [19]. Dispersion up to third order is considered, as higher-order terms are negligible. The interference of two identical time-stretched pulses with a delay of $\Delta\tau$ provides information on the difference in instantaneous wavenumbers:

$$\Delta\nu(t) = \nu(t - \Delta\tau) - \nu(t) = \frac{1}{2\pi c}\frac{\Delta\tau}{\phi_2}\left\{1 - \frac{\phi_3 t}{\phi_2^2}\left(1 - \frac{\Delta\tau}{2t}\right)\right\}. \tag{2}$$

As described earlier, the measured $\Delta\tau$ was 93.5 ps. A curve fitting for the measured chirp rate (Fig. 2(c)) was performed using Equation (2) within a range of ±5 ns around the center of the pulse. The evaluated values were $\phi_2$ = 5113.0 ± 0.4 ps$^2$ and $\phi_3$ = -31.7 ± 0.7 ps$^3$. These values are in excellent agreement with the theoretical estimates of $\phi_2$ = 5109 ps$^2$ and $\phi_3$ = -32 ps$^3$ for a 20-km-long DCF, based on the manufacturer-specified GVD and TOD coefficients of 255.5 ps$^2$/km and -1.6 ps$^3$/km, respectively.

Next, we conducted high-resolution MIR spectroscopy of methane gas using the developed system. The sample was a 5-cm-long methane gas cell (CH$_4$-T (25×5)-10-MgF$_2$, Wavelength References) at a pressure of 10 Torr. Figures 3(a) and 3(b) show the continuous temporal waveforms of MIR pulses with methane gas absorption measured at a rate of 50 MSpectra/s and their 100-times averaged waveform, respectively. The SNR of a single-shot time-stretched pulse without a sample was 14, calculated as the ratio of the peak intensity of the waveform, smoothed by a Savitzky-Golay (SG) filter [20], to the standard deviation of residuals obtained by subtracting the smoothed curve from the original waveform. Dispersive profiles on the absorption lines were observed, attributed to an insufficient chirp relative to the decay time of molecular vibrations, an effect referred to as the near-field propagation effect [21]. This phenomenon is analogous to the spatial near-field (Fresnel) diffraction pattern based on time-space duality. The small ripples around 15 ns in Fig. 3(b) are due to electrical reflections caused by a slight impedance mismatch in the detection system. Figure 3(c) compares the 100-times averaged spectrum with a simulated spectrum calculated using the

HITRAN database. The displayed spectrum was normalized to the baseline measurement without the sample. A detailed theoretical description of the simulation can be found in previous work [21].

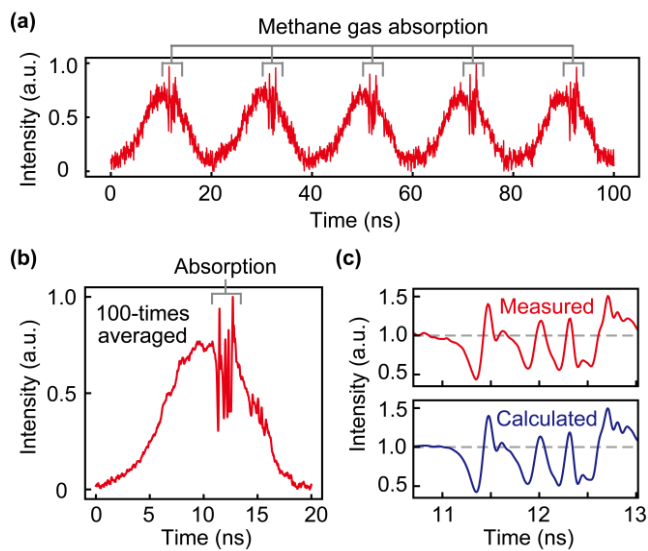

Fig. 3. MIR absorption spectroscopy of methane gas (a) Continuously measured time-stretched MIR pulses at a rate of 50 MSpectra/s. (b) Temporal waveform averaged over 100 measurements. (c) Comparison of dispersive absorption lines in the measured spectrum and calculated spectrum based on the HITRAN database.

The transmittance spectrum was retrieved from the measured spectrum using an iterative gradient-descent (GD) algorithm [21,22]. Figure 4 presents the transmittance spectrum recovered from the 100-times averaged temporal waveform shown in Fig. 3(b). The spectral resolution was 0.086 cm$^{-1}$ (2.6 GHz), determined by the full width at half maximum of the detection system's impulse response, measured as 84.1 ps using unstretched MIR pulses. The number of spectral elements was 220 over the range of 19.0 cm$^{-1}$. For comparison, a reference spectrum calculated based on the HITRAN database with the same resolution is also displayed in Fig. 4, demonstrating good agreement with the measured spectrum.

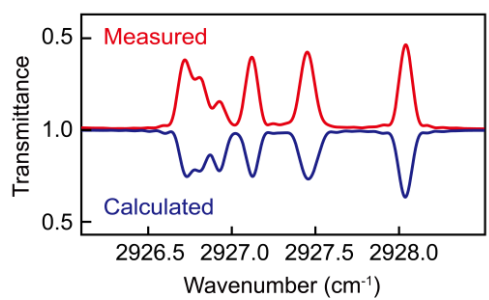

Fig. 4. Transmittance spectrum of methane gas retrieved from the 100-times averaged temporal waveform shown in Fig. 3(b) using the GD algorithm (red) and a reference spectrum calculated from the HITRAN database at a spectral resolution

of 0.086 cm$^{-1}$ (blue).

**Discussion**

The performance of our time-stretch frequency-swept laser critically depends on the use of low-loss optical fibers. A straightforward approach to stretching MIR pulses is to use optical fibers transparent in the MIR region, such as chalcogenide, fluoride, ZBLAN, or hollow core fibers. However, these fibers exhibit significantly higher losses compared to those used in the NIR region. For instance, time-stretching with an MIR fiber would result in losses on the order of tens to hundreds of dB/km [23]. Our down-conversion approach effectively addresses this challenge by leveraging low-loss NIR fibers, enabling efficient time-stretching without the high attenuation characteristic of MIR fibers.

Our frequency-swept MIR laser has the potential for improvement in several directions. First, the sweep rate could be increased by employing a mode-locked laser at a higher repetition rate. However, this would reduce spectral resolution unless paired with a higher-bandwidth detection system. Second, the spectral sweep range is currently limited by the phase-matching conditions of the waveguide PPLN crystal. A shorter nonlinear crystal could broaden the spectral range but would reduce conversion efficiency. Alternatively, incorporating multiple CW lasers with different wavelengths in the DFG process could expand the spectral range, allowing efficient access to discrete gas-phase absorption bands sparsely distributed across the MIR region. Third, adopting a fiber-coupled PPLN waveguide would make our system fully fiber-connected, resulting in a compact and robust laser setup. Finally, the detection efficiency, currently limited by the QCD's sensitivity (a few mA/W), could be improved by implementing an up-conversion detection system. Using an NIR photodetector with a sensitivity of a few A/W could enhance the SNR, even when accounting for the conversion efficiency of the up-conversion process.

It is valuable to compare the advantages and disadvantages of the down-conversion-based technique developed in this work and the previously reported up-conversion-based technique (UC-TSIR) [21] for MIR spectroscopy applications. The primary difference lies in the order of the sample and the stretching fiber: in the down-conversion technique, the pulses are stretched before the sample, whereas in the up-conversion technique, the sample interaction precedes the stretching process. The down-conversion approach avoids undesired nonlinear effects on the sample, making it suitable for linear spectroscopy applications. Conversely, the up-conversion technique is advantageous for, for example, time-resolved pump-probe spectroscopy, where sample interaction with ultrashort pulses is required. Additionally, the coupling efficiency to the stretching fiber is typically higher in the down-conversion method, as scattering from the sample in the up-conversion system can degrade the spatial mode, reducing efficiency. Another critical distinction is the complexity of the laser systems required. The down-conversion technique uses an off-the-shelf mode-locked fiber laser and a CW laser, providing a compact and practical solution. In contrast, the up-conversion technique requires a bulky and complex MIR femtosecond laser source, such as a femtosecond optical parametric oscillator, which increases the system's size and complexity.

## Conclusion

In conclusion, we developed a MIR frequency-swept source operating at a scan rate of 50 MScans/s, utilizing a 1.5-μm fiber time-stretcher. With a spectral bandwidth of 19.0 cm$^{-1}$, the system demonstrated high-speed MIR spectroscopy of methane gas, achieving a spectral resolution of 0.086 cm$^{-1}$ (220 spectral elements) at a measurement rate of 50 MSpectra/s. This rapid, compact, and robust MIR frequency-swept laser shows significant potential for field applications, including on-site combustion diagnosis via MIR absorption spectroscopy and factory inspections using MIR-OCT [24].

## Funding

Japan Society for the Promotion of Science (23H00273), JST FOREST Program (JPMJFR236C).

## Acknowledgements

We thank Zicong Xu for providing constructive comments on the manuscript.

## Disclosures

T.K., T.N., K.H., and T.I. are the inventors of a filed patent on the frequency-conversion time-stretch spectroscopy.

## Data availability

The data presented in the manuscript are available from the corresponding author upon reasonable request.